# Pulsed laser deposition with simultaneous *in situ* real-time monitoring of optical spectroscopic ellipsometry and reflection high-energy electron diffraction

J. H. Gruenewald, J. Nichols, and S. S. A. Seo[a]

*Department of Physics and Astronomy, University of Kentucky, Lexington, KY 40506*

We present a pulsed laser deposition (PLD) system that can monitor growth by simultaneously using *in situ* optical spectroscopic ellipsometry (SE) and reflection high-energy electron diffraction (RHEED). The RHEED precisely monitors the number of thin-film layers and surface structure during the deposition and the SE measures the optical spectra of the samples simultaneously. The thin-film thickness information obtained from RHEED facilitates the SE modeling process, which allows extracting the *in situ* optical spectra, i.e. the dielectric functions, of thin-films during growth. The *in situ* dielectric functions contain indispensable information about the *electronic* structure of thin-films. We demonstrate the performance of this system by growing $LaMnO_{3+\delta}$ (LMO) thin-films on $SrTiO_3$ (001) substrates. By using *in situ* SE and RHEED simultaneously, we show that real-time thickness and dielectric functions of the LMO thin-films can be effectively extracted. The simultaneous monitoring of both optical SE and RHEED offers important clues to understand the growth mechanism of atomic-scale thin-films.

---

[a] E-mail: a.seo@uky.edu

**INTRODUCTION**

Pulsed laser deposition (PLD) is a widely used method for growth of various thin-films, superlattices, and heterostructures.[1,2] It has been used extensively over the past two decades in the development of thin-films which exhibit interesting physical properties such as strongly-correlated electronic materials,[3] high-temperature superconductors,[4] colossal magnetoresistance materials,[2] ferroelectrics,[5] compound semiconductors, and nanocrystalline materials.[6]

PLD has remained popular for several reasons: it features the stoichiometric transfer of the target material to the film; it is compatible with both low and high vacuum pressure environments ($\leq 0.5$ Torr); and multiple materials can be easily added to a system for heterostructure synthesis. Additionally, atomic-scale oxide thin-films and heterostructures grown by PLD are shown to be comparable to samples grown by Molecular Beam Epitaxy (MBE).[7] Reflection high-energy electron diffraction (RHEED) has been successfully used to monitor the film growth in real-time by measuring the structural reconstruction and number of unit-cell layers.[8,9] However, the requirement of low-pressure operation and its inability to assess a material's electronic structure are a few shortcomings of RHEED. While advanced multiple-stage RHEED has overcome the first problem by increasing the maximum operational oxygen partial pressure to 0.5 Torr,[10, 11] there remains a huge demand for *in situ* monitoring of the electronic structure of a material during growth. Optical spectroscopic ellipsometry (SE), which is operational regardless of partial pressure, has been used as an alternative method to monitor a sample's electronic structure in real-time.[11, 12] However, SE is not used as frequently as RHEED since a complicated modeling process is required to analyze the SE data. Moreover, when *in situ* SE is performed without any additional supporting measurements, the results of the modeling process are difficult to validate due to multiple unknowns, e. g. fit parameters such as thin-film thickness and complex optical constants.

Here we demonstrate that the simultaneous use of both RHEED and SE in a customized PLD system can yield indispensable information about the growth mechanism of thin-films by probing both their ionic and electronic structure in real-time. The growth mode, number of layers, and diffraction patterns obtained by the RHEED data can be used in choosing an appropriate model to apply to the *in situ* SE spectra. To check the validity of the model, we show that the RHEED and SE have yielded the same growth rate of our test thin-films described below. The *in situ* SE allows for fast, real-time monitoring of the optical constants and electronic properties of the sample during synthesis. The simultaneous *in situ* monitoring techniques not only give us a better understanding of the growth process but also allow us to have better control in optimizing the growth parameters.

## I. SYSTEM OVERVIEW

Our PLD system with dual *in situ* SE and RHEED components is schematically drawn in Figure 1. The vacuum chamber is custom-made (Rocky Mountain Vacuum Tech.) in order to optimize data acquisition and to enable simultaneous real-time *in situ* measurement of RHEED and Spectroscopic Ellipsometry (SE). There are six ports specifically designed and implemented into this system in order to perform dual *in situ* RHEED and SE measurements. Two of these ports are used by the RHEED electron gun and detector (Phosphor screen and CCD camera assembly). They are positioned such that the electron beam is incident at a grazing angle to a substrate (sample) surface inside the chamber. The other four ports are positioned such that *in situ* ellipsometry can be performed simultaneously with RHEED at the angles of either 65° (not pictured in Fig. 1) or 75° to the substrate's surface normal direction. There is also an additional antireflection-coated quartz viewport for the pulsed excimer laser and is positioned such that focused laser beam pulses are incident to the target at an angle of 45°. While conventional PLD systems have been designed to perform either *in situ* RHEED or SE, our system is optimized for the simultaneous monitoring of both RHEED and SE, which better equips us to interpret the *in situ* data as discussed below.

The RHEED consists of an electron gun (Staib 30 kV model), a phosphor screen, and a charge-coupled device (CCD) camera (shielded from external light by a screen) attached to the PLD chamber. The electron beam is incident to the surface of the substrate at a grazing angle (≤ 3°) and the reflected electron beam is collected by the phosphor screen and is measured by the CCD camera.

The SE (J. A. Woollam M-2000-210 *in situ* model) consists of light source and analyzer assemblies mounted to the PLD chamber. White light from a broadband Xenon arc lamp is polarized in the SE source assembly by a linear polarizer and compensator. The light beam reflects off the sample and is directed to a rotating analyzer in the SE analyzer assembly, which is sent to a spectrometer (monochromator)/CCD detector through an optical fiber. In our system, the SE source and analyzer assemblies can be mounted to the PLD chamber with an incident angle of either 65° or 75° relative to the sample's normal axis, which are close to the Brewster angles and are good for characterizing many oxide thin-films.[13] The SE can take real-time dynamic scans every 25 ms in the spectral range of 210 – 1000 nm in wavelength (1.2 – 5.9 eV in photon energy). The spectral range of the SE is chosen for oxide materials, which allows for the monitoring of optical transitions between the correlated electron bands (Hubbard bands) and the charge-transfer transitions between the O 2$p$ band and transition-metal $d$ bands.

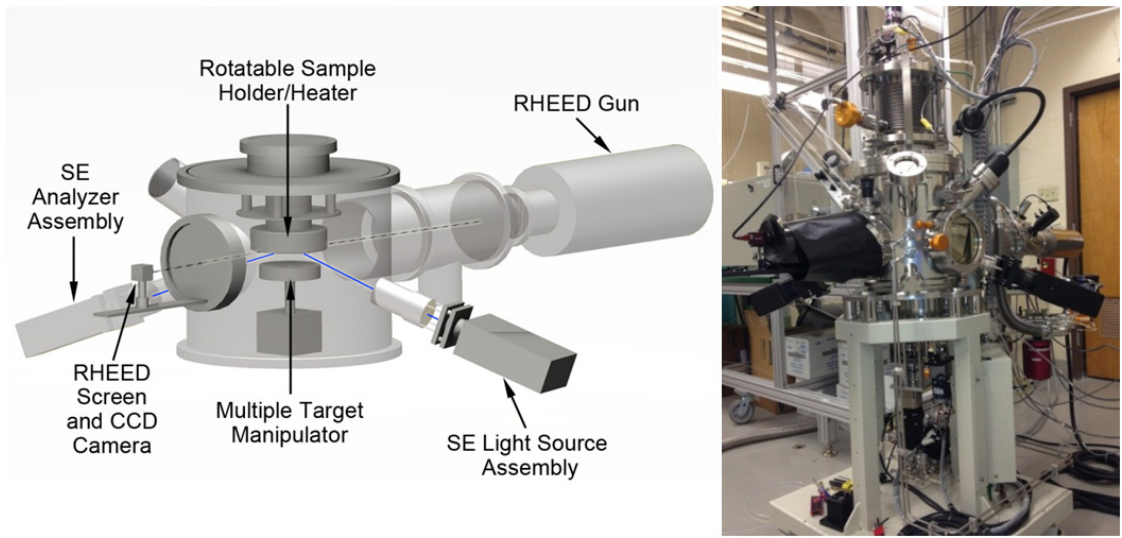

FIG. 1. (Color online) Schematic drawing (left) and photograph (right) of the PLD system with dual *in situ* RHEED and spectroscopic ellipsometry components taking simultaneous measurements.

In order to enable the simultaneous monitoring, the substrate holder (heater) is made rotatable with an adjustable (both tilting and height) z-axis stage so that the initial specular reflection of both RHEED and SE can be maximized. During the growth of a thin-film, the RHEED and SE data are simultaneously and continuously taken (maximum data acquisition rate of up to 40 Hz) without the need of suspending growth. The chamber is designed to enable sample alignment with the electron and photon beams by adjusting both the height and azimuthal angle of the sample holder. This is essential to perform both RHEED and SE measurements simultaneously during the entire growth process (Fig. 1). Below we discuss in detail how this system performs by testing the growth of epitaxial LaMnO$_{3+\delta}$ (LMO) thin-films on single crystalline SrTiO$_3$ (STO) substrates (001), whose compounds have drawn considerable attention due to interesting orbital characteristics[14] and the strong colossal magnetoresistance effect.[15]

## II. REFLECTION HIGH-ENERGY ELECTRON DIFFRACTION

Figures 2 (a)-(c) and (d) show the RHEED diffraction patterns and intensity oscillations taken during the growth of an LMO thin-film on an STO substrate, respectively. At the initiation of the deposition process, the electron beam is incident at a grazing angle to an atomically flat STO substrate prepared using the method described in Ref. 16. The oscillations of the intensity curve in Fig. 2 (d) show that LMO unit-cell layers are being deposited on the STO substrate, but the monotonic decay of the intensity curve indicates island growth is also occurring (layer-by-layer + island growth mode).[9,17] In addition, this decay of the intensity is typically seen when there is a lattice mismatch between the thin-film and the substrate.[18] However, the time scale of the decay is much larger than that of the oscillations, so a reasonably smooth epitaxial growth of the LMO thin-film can be deduced from the RHEED intensity pattern. Additional *ex situ* measurements, such as atomic force microscopy (AFM) are also important in final thin-film characterization and aids in validating the measurements made by the *in situ* SE and RHEED. According to the *ex situ* AFM topographic images of the sample before and after the growth, the thin-film still retains the step terraces of the substrate (see Figs. 2 (g) and 2 (h)). Hence, the thin-film's surface roughness is on the atomic scale and can be neglected in SE modeling even though the growth is not a perfect layer-by-layer growth. The RHEED patterns and intensity oscillations can be used to monitor the film's surface reconstruction and the number of unit-cell layers, respectively. These RHEED measurements of the thin-film deposition give indispensable information to validate the model used in extracting the optical constants from the SE spectra, as described in Section IV below. In our test growth, we observed clear RHEED intensity oscillations during the first 400 s of the deposition, corresponding to the LMO thin-film having a thickness of roughly 6 nm (15 unit-cells of LMO).

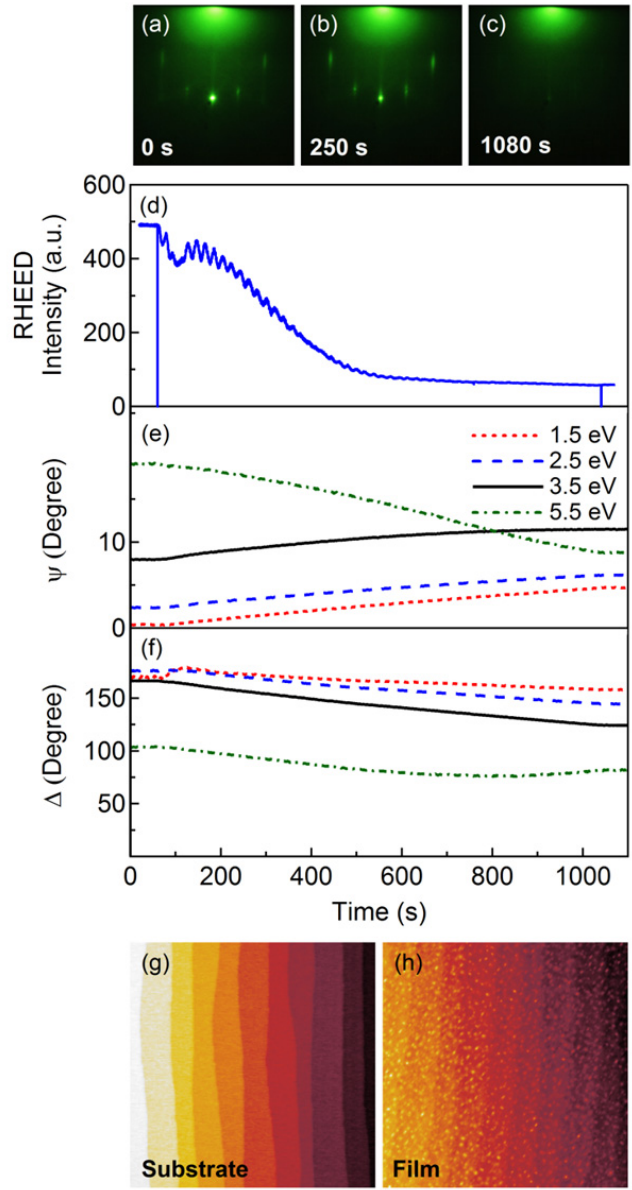

FIG. 2. (Color online) (a)-(f) *In situ* data taken simultaneously by RHEED and SE during LMO thin-film growth on STO substrate. RHEED diffraction patterns taken at times (a) 0 s, (b) 250 s, and (c) 1080 s. (d) RHEED intensity oscillations. The two vertical lines indicate the start and finish of growth. SE dynamical spectra taken of (e) Ψ and (f) Δ at 0.5 Hz during the thin-film growth. Surface topology of (g) STO substrate and (h) LMO thin-film taken *ex situ* by AFM. Growth Conditions: $P_{O2}$ = 30 mTorr, Substrate temperature $T_s$ = 700°C.

## III. OPTICAL SPECTROSCOPIC ELLIPSOMETRY

Polarized light from the *in situ* SE is incident upon the sample, and the two polarization states parallel (p) and perpendicular (s) to the plane of incidence are each changed separately according to the complex Fresnel reflection and transmission coefficients $\tilde{R}_i$ and $\tilde{T}_i$, where $i = p, s$ for $p$-polarized and $s$-polarized light, respectively.[19] These coefficients directly determine each reflected polarization's attenuation $|R_i|$ and phase change $\delta_i$. The light beam is then sent through the SE's rotating analyzer and the resulting intensity modulation is measured by a CCD array detector. The modulated signal can be fit to a Fourier series whose first two coefficients are determined by the Hadamard method. These coefficients, along with the initial *p*- and *s*- polarization state intensities, are used to find the relative attenuation $|\tilde{R}_p|/|\tilde{R}_s|$ and phase change shift $\delta_p - \delta_s$ of the reflected light beam. In ellipsometry, it is convenient to define

$$\tan \Psi \equiv \frac{|\tilde{R}_p|}{|\tilde{R}_s|} \tag{1}$$

$$\Delta \equiv \delta_p - \delta_s \tag{2}$$

$$\rho \equiv \tilde{R}_p / \tilde{R}_s = \tan \Psi \, e^{i\Delta}. \tag{3}$$

The angles $\Psi$ and $\Delta$ are the ellipsometer angles measured by the SE and represent the differential changes in amplitude and phase experienced upon reflection by the two *p*- and *s*- polarization components. The $\tan \Psi$ and $\cos \Delta$ are measured for the full spectral range of the light source (1.2 – 5.9 eV). During the growth of a thin-film on a substrate, the *in situ* SE can take dynamic full spectra at 40 Hz. Figures 2 (e) and 2 (f) show the real-time evolution of $\Psi$ and $\Delta$ at four arbitrary photon energies as measured by our *in situ* SE for the same LMO thin-film grown on an STO substrate as discussed above. Figure 3 shows full SE spectral scans of this growth at times 0 s, 500 s, and 1090 s. Since our test growth of an LMO thin-film on an STO substrate is slow, we are able to achieve reasonably good data by choosing an acquisition rate of 0.5 Hz. Using these measured values of Ψ and Δ, the real-time optical constants, dielectric functions, and the LMO film thickness can be obtained through a modeling process described in the following section.

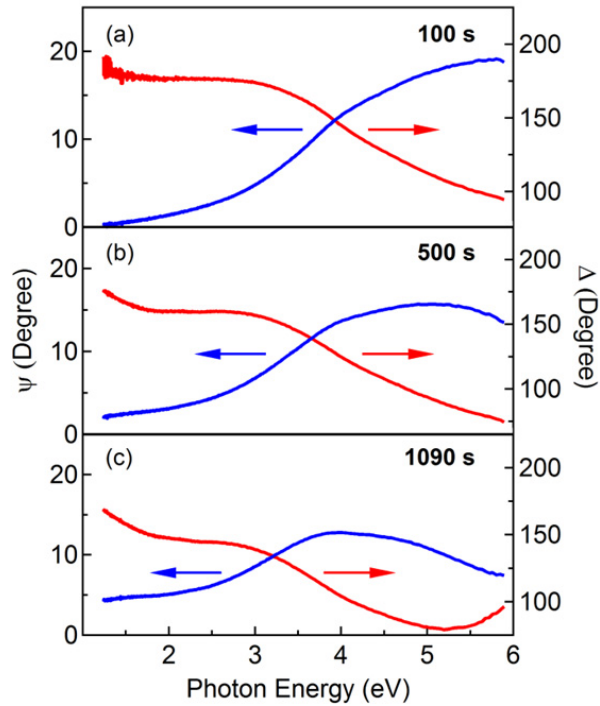

FIG. 3. (Color online) Full SE spectral scans taken from 1.2 eV to 5.9 eV are shown at times (a) 0 s, (b) 500 s, and (c) 1090 s during the LMO film deposition.

## IV. THICKNESS AND OPTICAL CONSTANTS OF THIN-FILMS

The growth rate of the thin-film can be independently monitored by both the RHEED and SE (see Fig. 4 (c)), acting as an experimental control for consistency between the two measurement devices. To find real-time film thickness and optical constants of the sample from the SE, the ellipsometry angles Ψ and Δ are used as input parameters to an appropriate sample model. The biggest advantage of simultaneous monitoring lies in the ability to use RHEED's diffraction and intensity patterns to choose and validate appropriate models which can be used in analyzing the *in situ* SE spectra. To date, the comprehensive model analysis required to interpret SE data has been a formidable task to accomplish in real-time measurements since it is rather difficult to obtain both thickness and optical parameters of thin-films using only SE data. Since we can obtain the information of a thin-film's growth mode and thickness by RHEED, we can bring more credibility to the SE spectral modeling process used for extracting the optical parameters from the SE spectra.

From Section II, the RHEED intensity curve (Fig. 2 (d)) and RHEED diffraction patterns (Figs. 2 (a)-(c)) indicate a 'layer-by-layer + island' growth of LMO on the STO substrate with surface roughness on the atomic scale. We confirmed these results with *ex situ* AFM as illustrated in Fig. 2 (h). These measurements reasonably justify modeling this sample as a thin, isotropic smooth layer (layer 1) on an isotropic substrate (layer 2) in an isotropic ambient (layer 0). When an appropriate model is applied, the optical constants and film thickness can be found from SE spectra. The complex Fresnel reflection coefficients for this three-phase system are found and are related to the SE spectra (Ψ and Δ) by Eq. 3 (see Appendix for details of calculation). From this relation, the thin-film's complex index of refraction $N_1$ ($N = n + ik$) and thickness can be numerically calculated at any time during the growth using proper initial values at each photon energy. Figures 4 (a) and 4 (b) show the distinguishable *n* and *k* spectra for the LMO thin-film and substrate both extracted at the end of the growth period.

Figure 4 (c) shows the real-time thickness of the LMO thin-film monitored by both SE and RHEED. From the RHEED intensity oscillations, the number of deposited LMO crystal layers is measured and the thickness is calculated by assuming each LMO unit cell has a thickness of ~3.95 Å.[20] Note that we used this thickness information acquired from RHEED to obtain the initial input parameters in our SE modeling process. The SE and RHEED thickness curves each yield consistent linear deposition rates of ~ 0.238 Å/s and ~ 0.208 Å/s, respectively. Therefore SE and RHEED collectively enable the overall film thickness to be determined in real-time during the entire growth process. Although the RHEED oscillations disappear after c.a. 400 s of growth, the same SE model is applied during the entire deposition. Both the SE and RHEED result in consistent thickness in the initial growth of the LMO thin-film, and SE can be used for thickness monitoring when the RHEED oscillations disappear, offering another advantage of simultaneous *in situ* monitoring. For example, if the LMO thin-film grown on STO described above was deposited in a chamber with only SE, there would be no knowledge of the film's *real-time* thickness during growth. Note that various *ex situ* characterizations are still very important but cannot provide *real-time* information during growth.

To obtain the electronic properties of the thin-film, the optical constants *n* and *k* can be converted into the complex dielectric function $\tilde{\varepsilon}(\varepsilon_1, \varepsilon_2)$ using the following relations:

$$\varepsilon_1 = n^2 - k^2 \tag{4}$$

$$\varepsilon_2 = 2nk, \tag{5}$$

Where $\varepsilon_1$ and $\varepsilon_2$ are the dispersion (real) and absorption (imaginary) components of the thin-film's complex dielectric function, respectively. The real-time dielectric functions can show the time evolution of bandgap energies and optical transitions during the entire deposition process. Figure 5 displays the complex dielectric function of the film at 100 s, 300 s, and 1090 s during the growth. The dispersion $\varepsilon_1$ and absorption $\varepsilon_2$ depend strongly on the thickness of the LMO thin-film. The *d-d* transitions at about 1.5 eV are initially suppressed during the growth. An appreciable spectral weight transfer in this narrow energy range of 1 eV – 6 eV is also apparent from Figure 5. At this moment, it is not clear what physical mechanism governs the thickness dependent spectral weight transfer, which is left for future study. The important point is that the simultaneous use of RHEED and SE enable us to model and extract the real-time dielectric functions of thin-films by reducing the number of unknown variables using the thickness information obtained from RHEED.

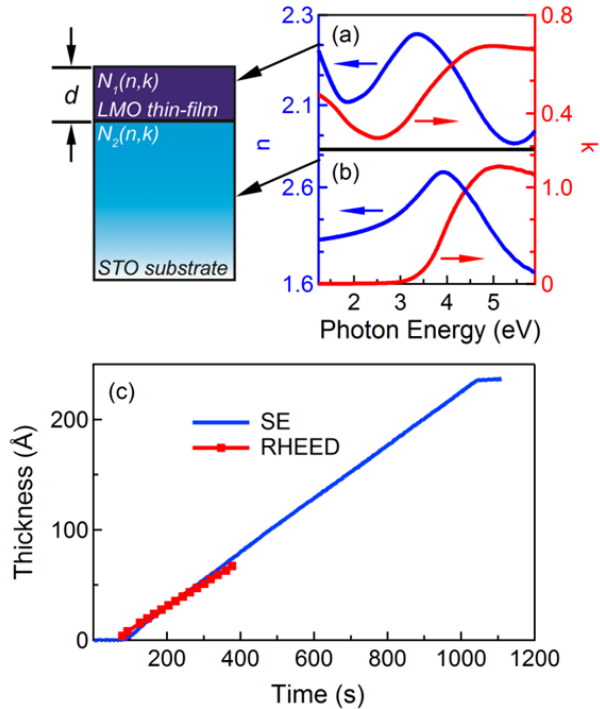

FIG. 4. (Color online) The optical constants from 1.2 – 5.9 eV for the (a) LMO thin-film $N_1(n,k)$ and (b) STO substrate $N_2(n,k)$ obtained from the *in situ* SE spectra at the end of the growth period. (c) Real-time thickness of LMO thin-film during the growth as determined independently by RHEED and SE. The measurements are self-consistent and support the validity of the model used to extract the optical constants and thickness from the SE spectra.

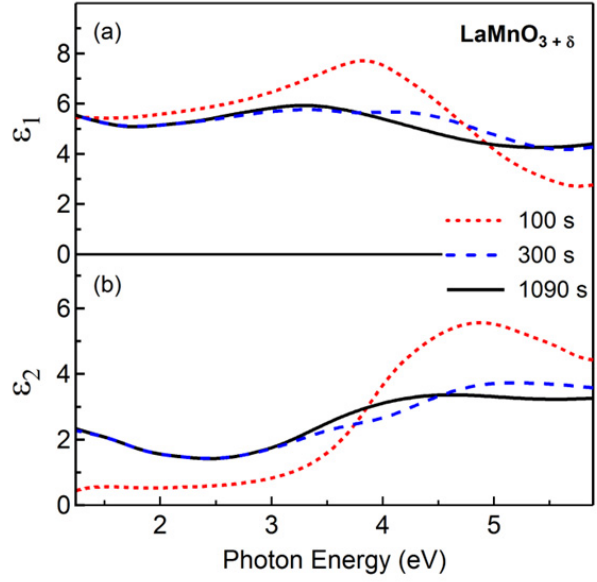

FIG. 5. (Color online) The (a) dispersion $\varepsilon_1$ and (b) absorption $\varepsilon_2$ coefficients as obtained from applying Eqs. 4 and 5 to the SE optical constants extracted at 100 s, 300 s, and 1090 s.

Note that the high temperature optical properties measured at the growth temperature (~ 700 ºC) are closely related to the thin-films' low temperature electronic properties and give us important information about their electronic structure.[21] For example, although band gap energies usually decrease with increasing temperature, they are material specific and are altered much more significantly by factors such as dopants or electronic structural change than by thermal effects. Note that, Kamaras *et al.*[22] demonstrated that depleting the oxygen content in $YBaCuO_{7-\delta}$ thin-films will change the films from a superconducting to a non-superconducting state; however, little temperature dependence was found in the optical spectra of these thin-films. Since *in situ* optical spectra are also sensitive to doping or thin-film thicknesses, we might be able to monitor important parameters such as oxygen deficiencies in oxide thin-films.[12] If there is a change in the dielectric functions (and bandgap energies) during growth, we need a rather sophisticated model to extract the optical parameters quantitatively. However, it is still possible to monitor the change qualitatively using a simple model. For example, in our growth of LMO, changes in the thickness dependence of the complex dielectric functions are observed even by using a simple thin, isotropic smooth layer-model (see Figure 5).

Our simultaneous real-time RHEED and SE technique can be effectively used for growing thin-films with targeted properties. For instance, tuning electronic properties such as band-gap energies can be done during the growth by directly monitoring the sample's optical spectra SE and adjusting the growth parameters during deposition. If a solid understanding of the relationship between the dielectric functions and the surface reconstruction of a thin-film layer is established, then the RHEED diffraction patterns will provide insight to choose a proper model for the SE data. These more elaborate real-time SE models are left for development and implementation in future work.

**CONCLUSIONS**

A PLD system with simultaneous *in situ* SE and RHEED is demonstrated for real-time monitoring of an oxide thin-film's surface structure, thickness, optical constants, and dielectric functions during the growth. When LMO thin-films are test grown on STO (001), the real-time optical constants and complex dielectric functions are obtained through a modeling process of the thin-film/substrate sample. The selected model works quite well since the thickness measurements between SE and RHEED are in good agreement. This monitoring technique is advantageous for several reasons. The SE spectral range is appropriately chosen to effectively probe various optical transitions of transition metal oxide thin-films, which give indispensable insight into the electronic band structure and its real-time change. The *in situ* measurements are non-destructive and are free from sample degradation due to atmospheric surface-interface layers. Fast, simultaneous measurements performed by the RHEED and SE allow the entire thin-film growth process to be observed and characterized. Finally, even though a model analysis is a formidable task to obtain the optical parameters and dielectric functions of thin-films, the simultaneously obtained RHEED data can validate a model for SE analysis, which gives us an important breakthrough in this field.

**ACKNOWLEDGEMENTS**

This research was supported by the NSF through Grant No. EPS-0814194 (the Center for Advanced Materials) and the Kentucky Science and Engineering Foundation with the Kentucky Science and Technology Corporation through Grant Agreement No. KSEF-148-502-12-303.

**APPENDIX**

For an isotropic, smooth layer (1) on an isotropic substrate (2) in an isotropic ambient (0), the complex Fresnel reflection coefficients are

$$\tilde{R}_p = \frac{r_{01p} + r_{12p} \exp[-i2\beta]}{1 + r_{01p} r_{12p} \exp[-i2\beta]} \qquad (A1)$$

$$\tilde{R}_s = \frac{r_{01s} + r_{12s} \exp[-i2\beta]}{1 + r_{01s} r_{12s} \exp[-i2\beta]}, \qquad (A2)$$

where $r_{01p}$, $r_{12p}$, and $r_{01s}$, $r_{12s}$ are the reflection coefficients at the 0-1 ambient-film and 1-2 film-substrate interfaces for *p*- and *s*- polarization states. Using appropriate boundary matching conditions from Maxwell's equations and application of Snell's law, these reflection coefficients can be found

$$r_{01p} = \frac{N_1^2 \cos \varphi_0 - N_0 N_1 \cos \varphi_1}{N_1^2 \cos \varphi_0 + N_0 N_1 \cos \varphi_1} \quad (A3)$$

$$r_{12p} = \frac{-N_1^2 \cos \varphi_2 + N_2 N_1 \cos \varphi_1}{N_1^2 \cos \varphi_0 + N_2 N_1 \cos \varphi_1} \quad (A4)$$

$$r_{01s} = \frac{N_0 \cos \varphi_0 - N_1 \cos \varphi_1}{N_0 \cos \varphi_0 + N_1 \cos \varphi_1} \quad (A5)$$

$$r_{12s} = \frac{-N_2 \cos \varphi_2 + N_1 \cos \varphi_1}{N_2 \cos \varphi_2 + N_1 \cos \varphi_1}, \quad (A6)$$

where $N_0$, $N_1$, and $N_2$ are the complex refractive indices of each respective layer ($N_j = n_j + ik_j$), $\varphi_0$ is the angle of incidence, and $\varphi_1$, $\varphi_2$ are the angles of refraction for the film and substrate, respectively.

$\beta$ is the phase change associated with light's propagation through the thin-film of thickness $d$ and is given for an isotropic film by

$$\beta = 2\pi \left(\frac{d}{\lambda}\right) (N_1^2 - N_0^2 \sin \varphi_0)^{1/2}. \quad (A7)$$

Substituting Equations A3-A7 into Equations A1 and A2, the complex Fresnel reflection coefficients can be related to the ellipsometric data ($\Psi$ and $\Delta$) by Equation 3.